\documentclass[aps,prb,preprint,groupedaddress]{revtex4-1}
\usepackage{graphicx}

\begin{document}


\title{
Role of the inner copper-oxide plane in interlayer Josephson effects in multi-layered cuprate superconductors
}


\author{Y. Nomura$^1$, R. Okamoto$^1$, T. Mizuno$^1$, S. Adachi$^2$, T. Watanabe$^2$, M. Suzuki$^1$, and I. Kakeya$^1$}
\affiliation{$^1$Department of Electronic Science and Engineering. Kyoto University, Kyoto, Japan\\
$^2$Graduate school of Science and Technology, Hirosaki University, Hirosaki, Japan
}
\date{\today}

\begin{abstract}
We find systematic signatures suggesting a different superconducting nature for a triple-layered cuprate Bi$_2$Sr$_2$Ca$_2$Cu$_3$O$_{10+\delta}$ with respect to a double-layer
through the properties of intrinsic Josephson junctions (IJJs).
Our measurements on the current-voltage characteristics reveal that the $c$-axis maximum Josephson current density is sensitive to the superfluid density in outer planes
while the critical temperature and the superconducting gap remain unaffected.
Switching dynamics of stacked IJJs exhibit that the fluctuation in gauge-invariant phase difference of an IJJ implies that the inner plane completely shields
the capacitive coupling between adjacent IJJs, which is essential for mono- and bilayered cuprates.

\end{abstract}

\maketitle

Cuprate superconductors have been widely debated for the past three decades.
Particulary, the underlying mechanism that determines the critical temperature ($T_c$) and methods to increase $T_c$ have received much attention.
An essential ingredient that determines $T_c$ is the number of CuO$_2$ planes ($n$) composing superconducting layers (SLs) that are separated by blocking layers (BLs).
It is known that $T_c$ increases with increasing $n$ for identical BLs in the crystal structure.
$T_c$ reaches its maximum value for $n=3$, beyond which it decreases with increasing $n$~\cite{Iyo2007}.
This is attributed to an insufficient carrier doping in inner planes (IPs) from BLs due to the presence of outer planes (OPs)~\cite{Trokiner1991,Kotegawa2001},
resulting in inequivalent superconducting gaps $\Delta$ of OP and IP~\cite{Ideta2010,Sekine2014}.

This inequivalent $\Delta$ arises issues on the relevance between $T_c$ and the maximum $c$-axis supercurrent $J_c$ in these trilayer materials
because the low-energy electrodynamics along the $c$-axis of cuprate superconductors are mostly described by the Josephson tunnelings between SLs~\cite{Kleiner1992,Tamasaku:1992,Matsuda1995,Koyama1996,Suzuki1998,Kakeya1998b}.
Moreover, in conventional superconductors, the maximum Josephson current is known to be determined by the superconducting gap of superconducting electrodes~\cite{Ambegaokar1963}.
However, in bilayer Bi$_2$Sr$_2$CaCu$_2$O$_{\rm 8+\delta}$ (Bi2212), the representative material of intrinsic Josephson junctions (IJJs),
the doping evolution of $J_c$ shows anti-correlation with that of $\Delta$~\cite{Suzuki2012a}.
Although doping evolutions of $\Delta$ and $J_c$ in trilayer Bi$_2$Sr$_2$Ca$_2$Cu$_3$O$_{\rm 10+\delta}$ have been reported,
the effect of the inequivalent $\Delta$ is smeared because of unwanted Bi2212 layers included in the samples~\cite{Yamada2003}.
The distribution of $\Delta$ in SLs are essential for intralayer electrodynamics, which have been observed in the infrared frequency regions in in YBa$_2$Cu$_3$O$_7$ (YBCO)~\cite{Tajima1993,Kojima2002}.
In multi-layered cuprates, multiple intralayer plasmons corresponding to different combinations between adjacent CuO$_2$ planes have also been observed~\cite{Hirata2012}.
Quite recently, Okamoto et al argue that the capacitive coupling~\cite{Koyama1996,Machida1999} plays an essential role for the intralayer Josephson coupling~\cite{Okamoto2016} in the context of the light-enhanced superconductivity~\cite{Hu2014} .

In this letter, we discuss the peculiarities of a trilayer cuprate superconductor in comparison to mono- and bilayer cuprates
through measurements of the Josephson critical current density ($J_c$), the superconducting gap ($\Delta$), and the Josephson switching rate ($\Gamma(I)$).
The measurements were performed on several mesa-structured samples of Bi2223 with a few IJJs of planar area $S=1.0 \mu{\rm m}^2$  as illustrated in Fig. \ref{fig:mesa}(a) and experimental details are described in Supplementary Materials~\cite{SM}.
We have found specific features for Bi2223 on the proximity effect of the triple-layered superconducting layer and the doping evolution of $J_c$.
The systematic difference between the phase-fluctuation temperature ($T_{\rm eff}$) extracted from $\Gamma(I)$ of the two topmost IJJs in Bi2223 compared to Bi2212 illuminates the role of capacitive coupling between adjacent IJJs~\cite{Koyama1996,Machida1999}.
These findings stimulate the development of a new intrinsic Josephson junction model
that includes electrodynamics between CuO$_2$ planes inside a superconducting layer of Bi-cuprates, which has been considered to be quenched.

\begin{figure}[tbp]
 \centering
\includegraphics[width = 0.9\linewidth]{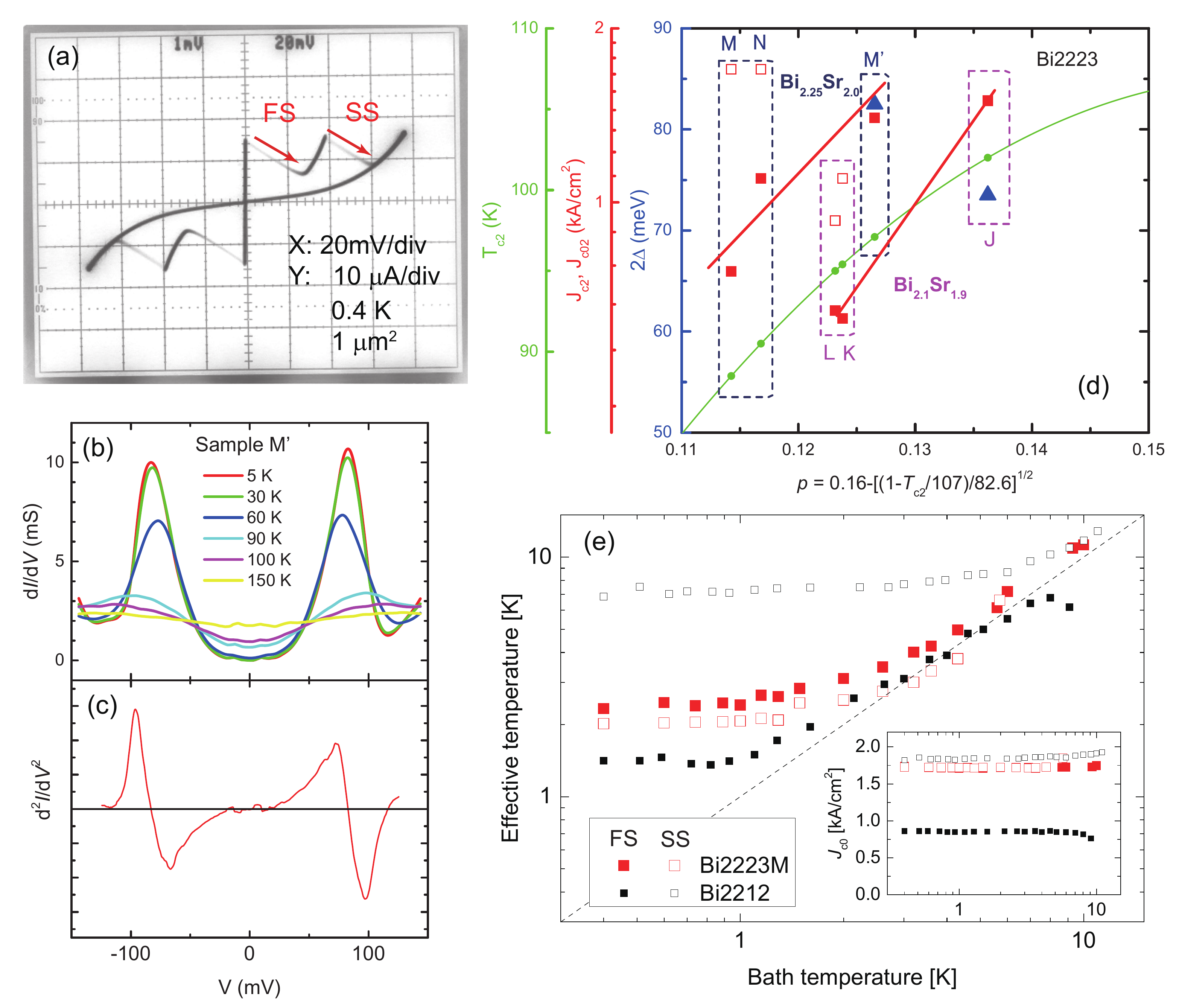}
 \caption{
	(a) IVCs of sample M at 0.4~K.
    IJJ1 and IJJ2 switch at FS and SS, respectively.
(b,c) $dI/dV$ spectra and the derivative (at 5 K) of mesa M'.
 $dI/dV$ spectra quantitatively corresponds to quasiparticle density of states thus the remarkable peaks below $T_c$ and minimum at $V=0$ are due to opening the superconducting gap and pseudogap.
 No signature implying the presence of different superconducting gap is seen.
 Small wiggling close to $V=0$ is due to incomplete suppression of Josephson current.
 (d) Doping evolution of mean value in critical current density of SS $J_{c2}$ (red squares), fluctuation-free critical current density of SS $J_{c02}$ (red open squares), $T_c$ (green circles), and superconducting gap $2\Delta$ (blue triangles) for the second IJJs of Bi2223.
 Two lines for different cation concentrations are drawn for $J_{c2}$ vs $p$.
(e) The relation between $T_{\rm eff}$ and $T_{\rm bath}$ for FS (solid) and SS (open) for sample M (red) and Bi2212 mesa C of Ref.~\cite{Nomura2017} (black).
In Bi2223, $T_{\rm eff}$ of the both switches is independent of temperature below 1~K
and $T_{\rm eff} \approx T_{\rm bath}$ holds above 2.0~K.
In Bi2212, $T_{\rm eff}$ for SS is much higher than that of FS at low temperature region.
In the inset, temperature dependence of fluctuation-free critical current densities $J_{\rm c0}$ are plotted with the same symbols.
\label{fig:results}
}
\end{figure}

First, we discuss a proximity effect for the triple CuO$_2$ planes found in the maximum Josephson current of the surface IJJ.
A current-voltage characteristic (IVC) of sample M at 0.4~K is shown in Fig.~\ref{fig:results}~(a).
The values of $J_c$ for first and second switches (labeled as FS and SS) of sample M are $J_{c1}=1.4$ and $J_{c2}=1.45$ kA/cm$^2$, respectively.
In Bi2212 mesa structures, $J_c$ of the first IJJ (IJJ1) is commonly reduced more than tens percent of the $J_c$ of other bulk IJJs (IJJ2)~\cite{SM,Nomura2017}.
This is due to the proximity effect to the normal electrode evaporated just above IJJ1~\cite{Gupta2004,Suzuki2008}.
However, in Bi2223, this reduction is significantly smeared.
We attribute this change of the reduction in $J_c$ to the presence of IPs.
In cuprate superconductors with the $d$-wave symmetry of superconducting gap, superfluid density $\rho_s$ determines both $T_c$~\cite{Uemura1991,Homes2004} and $J_c$
because carrier doping significantly influences to the extent of the Fermi arc in the $k$-space and
consequently alters superconducting properties even for cases with an identical superconducting gap~\cite{Kanigel2006,Lee:2007Nature,Yoshida2006,Kambara2013}.
Here, SLs (double and triple CuO$_2$ planes for Bi2212 and Bi2223, respectively) are numbered from the top to the bottom as SL1, SL2, etc.,
and CuO$_2$ planes are labeled as UOP1, IP1, and LOP1 for the case of SL1 in Bi2223, as shown in Fig. \ref{fig:mesa} (a,b).
BLs are also numbered as BL1, BL2, etc. similar to SLs.
In Bi2212, as depicted in Fig. \ref{fig:mesa} (c), Cooper pairs in UOP1 (closest to the normal metal electrode) mainly contribute to the diffusion and Cooper pairs in LOP1 are
less but considerably diffused.
As a result, the penetration of $\rho_s$ into underlying BLs is significantly reduced.
In Bi2223, IP1 plays the role of the LOP1 in Bi2212.
Thus $\rho_s$ in the third CuO$_2$ plane (which is LOP1) is much less influenced.
Therefore, the $J_c$ of a multilayered IJJ is determined by a combination of $\rho_s$ in a pair of OPs sandwiching a BL (viz. LOP1--UOP2 and LOP2--UOP3 for $J_{c1}$ and $J_{c2}$, respectively).

Second, we exhibit that chemical disorder of the Sr site significantly reduces $\rho_s$ of OPs irrespective to $T_c$.
Figure \ref{fig:results}~(d) shows $J_{c2}$ as a function of doping ($p$) estimated from bulk critical temperature $T_{c2}$ through an empirical formula~\cite{Tallon1995} with
an assumption of the optimum $T_{c2}$ of 107 K~\cite{Yamada2003}.
Although the relationship is relied on mono- and bilayer cuprates, the formula is valid to express relative difference in doping of underdoped trilayer cuprates~\cite{Jover1996}.
It is clearly seen that the mean $J_{c2}$ of samples K and L is considerably lower than samples of M and N despite their higher $T_c$.
One can infer two exponential doping evolutions in $J_c$ corresponding to the difference in Bi/Sr chemical compositions of crystals.
This is in sharp contrast to the case of Bi2212, where the doping evolution of $J_c$ falls into a single exponential dependence
irrespective of the difference in Bi/Sr ratios~\cite{Suzuki2012a}.
It is known that Bi2212 and Bi2201 crystals with Sr concentration below 2.0 tend to have a lower optimum $T_c$ than perfectly stoichiometric crystals.
The disorder in CuO$_2$ planes, induced by the partial substitution of Bi at the Sr sites adjacent to the CuO$_2$ plane, reduces $T_c$~\cite{Eisaki2004}.
In the present study on Bi2223, OPs of batch A (samples J, K, and L) are more disordered than OPs of batch B (samples M, M', and N) because more Sr are substituted for Bi in batch A~\cite{SM}.
We consider this as a reason for lower $J_{c}$ of mesas K and L compared to mesas M and N with an anti-correlation between $J_{c}$ and $T_c$.
The present results, nevertheless, show that $T_c$ of Bi2223 is not likely governed by the disorder of the CuO$_2$ planes.
Although no complete doping evolutions of $J_c$ and $T_c$ have been clarified in this work,
a simple comparison between batches A and B implies that $J_c$ is determined by both the doping and disorder of OP
whereas $T_c$ is determined by doping of IP.

The disorder of the Sr site does not affect to the magnitude of the superconducting gap.
We have performed intrinsic tunneling spectroscopy in samples J and M'.
Quasiparticle $dI/dV$ vs $V$ spectra of sample M' are shown in Fig. \ref{fig:results} (b).
The obtained $2\Delta$ = 83 meV at $T= 5$ K of the sample M' is roughly consistent with the published $2\Delta$ = 82 meV by Yamada et al.~\cite{Yamada2003} for $T_c=$ 96 K in the $10 \times 10\mu{\rm m}^2$ and $N \sim 10$ mesa made of a crystal with identical nominal composition to batch A and slight Bi2212 intergrowth impurities.
The obtained $\Delta$ is close to the smaller superconducting gap of 35 meV obtained by point contact measurements for a crystal with $T_c=101$ K~\cite{Sekine2014}.
Comparing with ARPES data from OP, our $\Delta$ is considerably smaller than the nodal gap $\sim$ 50 meV~\cite{Ideta2010}.
Furthermore, the proximity effect from the Ag electrode is not presumed to reduce the amplitude of superconducting gap of UOP1.
Quasiparticle tunneling spectra between superconductors with different superconducting gaps $\Delta_1$ and $\Delta_2$ exhibit
two $dI/dV$ peaks at $eV=\Delta_1+\Delta_2$ and $|\Delta_1-\Delta_2|$ ~\cite{Wolf2012}.
In the present study, however, such a signature is not found even for the second-order derivative $d^2I/dV^2$, as shown in Fig. \ref{fig:results} (c).
Thus we conclude that the proximity effect reduces carrier concentrations of CuO$_2$ planes of SL1 and results in the decrease in $J_{c1}$.
Spectroscopic measurements such as ARPES and STM to determine superconducting gap and carrier density in the momentum space will validate our arguments.

The difference of intralayer electrodynamics between bi- and trilayer cuprates is illuminated
by fluctuation measurements of the Josephson switching of an individual IJJ more explicitly in terms of the capacitive coupling between adjacent IJJs
because its length scale is the order of 1 nm in BSCCOs~\cite{Koyama1996}.
The Josephson switching rate $\Gamma(I)$ consists of quantum ($\Gamma_{\rm Q} \propto \exp[-\omega_p^{-1}]$) and thermal ($\Gamma_{\rm T} \propto \exp[-T^{-1}]$) contributions, where $\omega_p$ and $T$ are the Josephson plasma frequency and temperature of the system, respectively~\cite{SM,Kramers,Caldeira1981}.
Figure~\ref{fig:results}~(e) shows phase-fluctuation temperature $T_{\rm eff}$ and fluctuation-free critical current density $J_{\rm c0}$ of FS and SS in sample M as functions of bath temperature $T_{\rm bath}$ together with the data in Bi2212~\cite{Nomura2017}.
$T_{\rm eff} - T_{\rm bath}$ behaviors for FS and SS are very similar in Bi2223 while $T_{\rm eff}$ for SS is much higher than $T_{\rm eff}$ for FS in the saturated low-temperature region $T_{\rm bath} < $ 1~K in Bi2212
despite of their similar $J_{c0}$ of SS as shown in the inset of Fig.~\ref{fig:results}~(e).
These are common among 4 Bi2223 mesas and 4 Bi2212 mesas with identical lateral size~\cite{Nomura2014,SM}.
Thus we suggest a naive conclusion that quantum fluctuation, $\omega_p$, of SS is not enhanced in trilayer cuprates in contrast to mono- and bilayer cuprates~\cite{Nomura2015,Nomura2017}.

This absence of the enhancement in $\omega_p$ is presumed to be attributed to a reduction of the capacitive coupling in Bi2223 through the following considerations.
It is noted that the possibility of heating as the origin of the increased $\omega_p$ for SS of Bi2201 and Bi2212 is certainly excluded by
a counterintuitive behavior in sample M that the saturated $T_{\rm eff}$ for SS ($T^*_2 =$2.0 K) is slightly lower than that for FS ($T^*_1=$ 2.3 K).
$T^*_2/T^*_1$ roughly corresponds to the enhancement factor of $\omega_p$.
In the presence of the capacitive coupling, IJJ2 couples with the Josephson oscillation $\omega_J=2eV/\hbar$ of IJJ1 on the verge of SS, where $V$ is the applied voltage to IJJ1.
Here, IVCs offer $\omega_J/2\pi \simeq$ 12 and 20 THz for Bi2212 and Bi2223, respectively.
Assuming an SL being a uniform superconducting electrode, IJJ1 directly couples to IJJ2.
Josephson coupling energy $E_J=\hbar J_{c0}S/2e$ of IJJ2 ($\sim$ 10 THz) is smaller than $\hbar \omega_J$ for both Bi2212 and Bi2223, thus the Josephson radiation results in excitations to continuous energy state out of the \emph{hollow} of tilted washboard potential.
This is an analogue of the photovoltaic effect.
The Josephson radiation to IJJ2 of Bi2223 damps to the factor of $0.3-0.5$ in comparison with the case of Bi2212 because of increase in screening superfluid carriers in SL2.

An introduction of the intralayer Josephson coupling explains the phenomena more microscopically.
The lowest intralayer Josephson plasmons lie at 14 and 15 THz for Bi2212~\cite{Zelezny2001} and Bi2223~\cite{Boris2002}, respectively.
The Josephson oscillation of IJJ1 may couple to the intralayer plasmons under a current bias.
In the bilayer case, the intralayer plasmon of SL2 directly couples to the Interlayer plasmon of IJJ2.
However in the trilayer case, the intralayer plasmon between UOP2 and IP2 is necessary to excite another intralayer plasmon between IP2 and LOP2 to couple to IJJ2.
This complicated process to couple between IJJ1 and IJJ2 is presumed to make the enhancement of the quantum fluctuation of IJJ2 invalid.
Thus we claim that the presence of IP provides significant difference of capacitive coupling in Bi2223 from those in Bi2201 and Bi2212.

Recent observations of the light-induced superconductivity in YBCO up to room temperatures~\cite{Hu2014} have been in investigated theoretically by Okamoto et al.~\cite{Okamoto2016}.
In their model, the capacitive coupling between the CuO$_2$ planes within a SL is taken into consideration on the basis of the alternating IJJ model ~\cite{Koyama2002}.
They propose that the moderate capacitive coupling between the bilayer of YBCO contribute to enhanced interlayer Josephson coupling.
However, in Bi2223, light-induced superconductivity at temperatures higher than its equilibrium $T_c$ has not been observed~\cite{Hu2017}.
This scenario is totaly consistent to our results described above.

In summary, the localized picture of superfluid density $\rho_s$ in CuO$_2$ planes explains the qualitative differences of the Josephson maximum currents and the superconducting gaps between Bi2212 and Bi2223.
The behavior of the phase-fluctuation revealed that the inner CuO$_2$ plane of Bi2223 shields the capacitive coupling between IJJs.
These specific features in Bi2223 propose a new model to describe the $c$-axis conductivity of multi-layered cuprates more precisely.

\begin{acknowledgments}
This work was partly supported by Grant-in-Aid for JSPS Fellows No. 16J11491.
We thank Osamu Sakai at University of Shiga Prefecture for allowing us to use the electron-beam lithography facility.

\end{acknowledgments}

\bibliography{mqt_2,library}
\begin{figure}[tbp]
 \includegraphics[width = 0.9\linewidth]{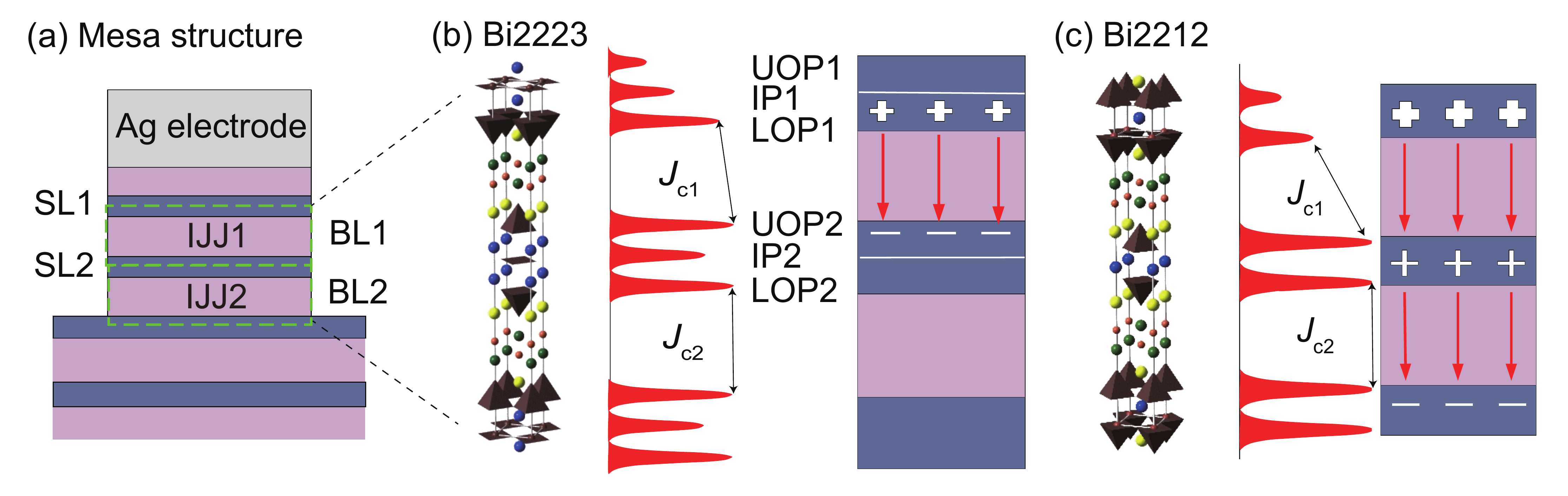}
\caption{
    (a) Cross-sectional schematic view of the mesa.
    (b) For Bi2223, crystal structure corresponding to the mesa cross section, sketch of distribution of superfluid density $\rho_s$ and schematic illustration of capacitive coupling from left to right.
    (c) Corresponding illustrations for Bi2212.
\label{fig:mesa}
}
\end{figure}

\end{document}